\documentclass[aps,pra,twocolumn,showpacs,floatfix]{revtex4}
\usepackage{graphicx}   

\begin{document}

\title{
Absorbing boundaries in numerical solutions of the time-dependent Schr\"odinger
equation on a grid using exterior complex scaling
}

\author{F. He, C. Ruiz, A. Becker}

\affiliation{
Max-Planck-Institut f\"ur Physik of komplexer Systeme,
N\"othnitzer Str.38, D-01187 Dresden, Germany
}

\date{\today}

\begin{abstract}
We study the suppression of reflections in the numerical
simulation of the time-dependent Schr\"odinger equation 
for strong-field problems on a grid
using exterior complex scaling (ECS) as absorbing boundary condition. 
It is shown that the ECS method can be applied in both the length and 
the velocity gauge as long as appropriate restrictions are applied in the ECS
transformation of the electron-field coupling.
It is found that the ECS method improves the
suppression of reflection as compared to the conventional masking
  technique in typical simulations of atoms exposed to an intense laser
  pulse. Finally, we demonstrate the advantage of the ECS
technique to avoid unphysical artifacts in the evaluation of high
harmonic spectra.
\end{abstract}


\maketitle

\section{Introduction}
With the progress in laser technology in recent years
\cite{Strickland85,Mourou98}, the focused laser field strengths
increased rapidly to exceed the strength of the Coulomb fields
that bind the electrons in the ground state of an atom or
molecule. Such intense light fields have led to the discovery of
novel aspects of light-matter interactions, such as multi-photon
ionization, above-threshold ionization, high harmonic generation,
multiple ionization to high charge states, Coulomb explosion  etc.
The lowest-order perturbation theory of light-matter interaction
is known to break down at light intensities above about $I \simeq
5\times 10^{12}$ W/cm$^2$ for near optical wavelengths
\cite{Faisal87}. Several nonperturbative theories of laser-matter
interaction, such as the numerical solution of the time-dependent
Schr\"odinger equation (TDSE) 
\cite{Kulander87,Krause92,Muller99,Nurhuda99}, ab-initio methods
based on the Floquet theorem 
\cite{Shakeshaft90,Burke91,Dimou92}, basic set expansion methods
 \cite{Tang90,Antoine95,Cormier96} or $S$-matrix and related
theories \cite{Keldysh64,Faisal73,Reiss80,Becker05}, have
been developed. Among these the solution of the TDSE on a
time-space grid is considered as a rigorous and powerful approach.
For the investigation of most of the above-mentioned intense-field
phenomena the simulation of an atom with $N$ electrons, which
would require the solution of a set of $3 \times N$ dimensional
partial differential equations, can be well approximated using the
single-active-electron (SAE) model \cite{Kulander87}. In the
latter only one of the electrons of the atom is considered to
become active and to interact with the external field, which
reduces the numerical problem to at most three dimensions. Any
symmetry of the external field may further reduce the
dimensionality.

Accurate solutions of time-dependent problems on the grid require
not only relatively dense grid points but, in general, also a huge
spatial extension of the grid to account for the release and the
motion of the electron sub-wave packets in the field. These
factors result in large memory and CPU requirements for the
numerical solution at high field intensities. Fortunately, the
important dynamics of several intense-field processes occur on a
spatial volume close to the atom or molecule. A few examples are
the determination of ionization rates \cite{Kulander87} and high
harmonic spectra \cite{Krause92} or the identification of dominant
pathways to single \cite{Harumiya02} or nonsequential double
ionization \cite{Baier06} of a molecule. These studies can be
therefore performed on a relatively small grid neglecting the
exact form of the outgoing ionizing parts of the wave function. It
however requires to suppress reflections from the edges of the
numerical grid, which can cause artificial effects, e.g. in the
form of spurious harmonics in high harmonic spectra
\cite{Krause92}.

In calculations based on the
numerical propagation of wave packets in different areas of physics and
chemistry
several techniques have been proposed to compensate for reflections,
including masking functions (or equivalently absorbing imaginary potentials)
\cite{Krause92,Santra02,Muga04},
repetitive projection and Siegert state expansion methods
\cite{Yoshida99}, complex coordinate rotation or exterior complex scaling
\cite{Ho83,McCurdy04} and others. In the numerical solution
of the time-dependent Schr\"odinger equation of atoms in intense laser fields
masking functions or absorbing potentials are the most
commonly used techniques
(e.g. \cite{Kosloff86a,Krause92,Ishikawa02,Ishikawa03,Santra04,Gordon06}).
The ECS technique is rarely used in this context up to now
\cite{McCurdy911,McCurdy912,McCurdy02}, which might be due to the fact, that
its application in the electric field (or length) gauge 
has been thought to be not fruitful \cite{McCurdy912}.

In this paper we re-examine the implementation of the
exterior complex scaling (ECS) method \cite{Nicolaides78,Simon79}
as absorbing boundary condition in simulations of 
strong-field problems.
In particular, we focus on the application of the ECS technique in 
  different gauges, namely the length and the velocity gauge. On the basis of
  results of simulations of 1D model atom exposed to an oscillating
  external field, we investigate appropriate restrictions of the ECS
  transformation of the external-field coupling to avoid undesired effects in
  the absorbing area.
Results for the probability density and momentum distributions 
for the interaction of the
hydrogen atom with an intense laser field are then compared with those
obtained using the standard masking function technique.
Finally, we consider a typical strong-field problem by using the ECS method to
calculate high harmonic spectra.

\section{Exterior complex scaling (ECS)
as absorbing boundary}

The complex scaling method has been widely used in physics and
chemistry (for a review, see \cite{Reinhardt82}), e.g. in the
theory of potential scattering \cite{Regge59}, calculation of
resonances in atoms and molecules \cite{Doolen74} or the
calculation of cross sections in scattering processes
\cite{Rescigno75}. According to this method the radial coordinate
of the particles are scaled by a complex phase factor, which
distorts the spectrum of the Hamiltonian such that the continuous
spectrum is rotated in the complex energy plane and the discrete
resonance eigenvalues are revealed. For our aim to introduce
absorbing boundaries at the edges of a numerical grid in
time-dependent simulations we make use of an extension of the
complex scaling method, namely the exterior complex scaling (ECS)
technique \cite{Simon79}, in which the spatial coordinates are
only scaled outside some distance from the origin.

\begin{figure}
\includegraphics[width=0.7\columnwidth]{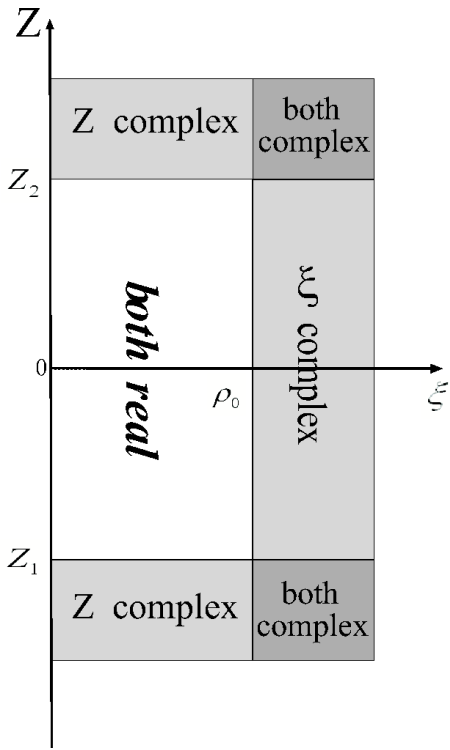}
\caption{\label{f1}
Scheme of the ECS coordinate transformation. Real coordinates
are used in the interior box, defined by $z_1\leq z\leq z_2$ and
$\rho \leq \rho_0$. The zones in gray mark the areas, where one or both
coordinates are complex.
}
\end{figure}

As discussed at the outset the ECS technique has been used before 
\cite{McCurdy911,McCurdy912,McCurdy02} in time-dependent studies of 
electron impact ionization as well as of the
motion of a charged particle in dc and ac electromagnetic fields. In the latter
context it has been mentioned \cite{McCurdy912} that it appears
to be not fruitful to apply ECS in the electric field (or
length) gauge. We now revisit this question and analyze below how
ECS can be used in both the length and the velocity gauges in simulations of 
atoms exposed to intense laser pulses.

\subsection{Implementation}

First, we illustrate the implementation of the ECS technique in 1D and 2D
time-dependent calculations, it is straightforward to
extend it to higher dimensions (c.f. \cite{McCurdy912,McCurdy04}. 
Let us consider
the non-relativistic electron dynamics 
in a time-independent (Coulomb) potential $V_0$ and an external field
governed by the
time-dependent Schr\"odinger equation in cylinder coordinates as
(Hartree atomic units, $m_e = \hbar = e = 1$, are used throughout):
\begin{equation}
\label{sch}
i{\partial \over \partial t}\psi(z,\rho;t)= [H_0(z,\rho)+V(t)]\psi(z,\rho;t)
\end{equation}
with the time-indepentent Hamiltonian
\begin{equation}
H_0(z,\rho) =
- {1 \over 2}{\partial^2\over \partial \rho^2}
- {1\over 2\rho}{\partial\over\partial\rho}
-{1\over 2}{\partial^2\over \partial z^2}+V_0(z,\rho)
\end{equation}
and time-dependent external-field coupling, given in length or
velocity gauge, as
\begin{eqnarray}
V(t) &=&
\left\{\begin{array}{ll}
zE(t),  & \ \ \textrm {length gauge}\\
{A\over c}\hat p_z,& \ \ \textrm {velocity gauge}
\end{array} \right.
\label{coupling}
\end{eqnarray}
Here $E(t)$ is the electric field and $A(t)$ is the vector potential of
  the external electromagnetic pulse linearly polarized along the
  $z$-direction.

The ECS transformation on the
two coordinates $z$ and $\rho$ can be given by
(c.f. Fig. \ref{f1}):
\begin{eqnarray}
\label{ECS_z}
Z &=&
\left\{\begin{array}{ll}
z_1+(z-z_1)\exp(i\eta)& \ \ \textrm {as $z<z_1$}\\
z& \ \ \textrm {as $z_1\leq z\leq z_2$}\\
z_2+(z-z_2)\exp(i\eta),&\ \ \textrm {as $z>z_2$}
\end{array} \right.
\\
\label{ECS_rho}
\xi &=&
\left\{ \begin{array}{ll}
\rho & \ \ \textrm{as $\rho \leq \rho_0$}\\
\rho_0+(\rho-\rho_0)\exp(i\eta),& \ \ \textrm{as $\rho>\rho_0$}
\end{array} \right.
\end{eqnarray}
where $\eta$ is the scaling angle with $0<\eta<\pi/2$. $z_1$,
$z_2$ and $\rho_0$ are labeled in Fig. \ref{f1} and define the
size of the interior box ($z_1\leq z\leq z_2$ and $\rho\leq\rho_0$)
within which both spatial coordinates are real. Outside (gray
zones in Fig. \ref{f1}) one or both coordinates are complex.

It is the aim of the ECS method to transform the outgoing wave
  into a 
function, which falls off exponentially outside the interior box, while the
wave-function keeps unchanged in the region where the coordinates are
real \cite{McCurdy04,Rescigno00,Moiseyev}. In case of the present problem of
an atom exposed to an oscillating linearly polarized electromagnetic pulse, we 
therefore 
investigate whether or not the transformed solution of Eq. \ref{sch} shows this
desired behavior. To this end and without loss of generality, 
we restrict our analysis to the $Z$ direction, i.e. the direction of the 
external field. The time-dependent solution of the 1D analogous of 
Eq. \ref{sch} in the complex area can be written as:
\begin{equation}
\psi(t+\Delta t) \sim
\exp[-i(H_0(Z)+V(t))\Delta t]\psi(t).
\label{solution}
\end{equation}
The time-independent operator in Eq. (\ref{solution}) is given by:
\begin{eqnarray}
\lefteqn{\exp[-iH_0(Z)\Delta t] =}
\nonumber\\
&&\exp\left[-i{\cos(2\eta)\over 2}{\hat p_z^2}\right]
\exp\left[-{\sin(2\eta)\over 2}{\hat p_z^2}\right]
\nonumber\\ 
&&\times\exp[-iRe(V_0)\Delta t]\exp[Im(V_0)\Delta t] 
\label{indep}
\end{eqnarray}
As discussed by McCurdy et al. \cite{McCurdy912} the exponent in the 
second factor on the right hand side of Eq. (\ref{indep}) is always negative
if $0<\eta<\pi/2$ and provides already the desired decay term. 
It is therefore important to note that the wave-function will be basically
absorbed in the complex area due to the transformed kinetic operator term 
as long as there are no 
counteracting effects from other terms in the Hamiltonian.
In general, it is therefore 
required that $Im(V_0) \le 0$ such that the last term in
Eq. (\ref{indep}) acts as an absorbing potential. 
In practice, 
the discontinuity in the real and imaginary part of the potential terms 
introduced by the complex scaling factor can generate some small reflections.  
An efficient way to avoid this numerical problem appears to keep the 
$Z$-coordinate in the potential term untransformed as a real number in the  
absorption area. 
Please note that it 
is unproblematic to abandon the additional absorption effect of the 
imaginary potential, since the decay of the wavefunction is already ensured via
the kinetic operator term.

The time-dependent potential in Eq. (\ref{solution}) can be written as
\begin{eqnarray}
\lefteqn{\exp[-iE(t)Z\Delta t]}
\nonumber\\
&& = \exp[-iE(t)(z_{1,2}+(z-z_{1,2})\exp(i\eta))\Delta t]
%\label{uns1}  
\nonumber\\
&& =
\exp[E(t)(z-z_{1,2})\sin\eta\Delta t]
\nonumber\\
&& \quad\times \exp[-iE(t)(z_{1,2}+(z-z_{1,2})\cos\eta)\Delta t]
\label{ecs1}
\end{eqnarray}
in the length gauge or as
\begin{eqnarray}
&&\exp\left [-i\left({A(t)\over c}\exp(-i\eta)\hat p_z\right)\Delta t \right ]
\nonumber\\
&& =
\exp\left[-\sin\eta {A(t)\over c}\hat p_z\Delta t\right]
\exp\left[-i\cos\eta {A(t)\over c}\hat p_z\Delta t\right] 
\label{ecs2}
\end{eqnarray}
in the velocity gauge. In the right sides of Eqs. (\ref{ecs1}) and 
(\ref{ecs2}), 
the second factors are oscillatory ones, and are simple scaled versions of the
external-field coupling as the wavefunction enters the complex area. 
The first factors, however, can act both as an absorber or as an undesired 
source, depending on the sign of the exponent. It is determined in the 
length gauge by the instantaneous sign of the oscillating electric field,
while in the velocity gauge, it equals the sign of the product $A(t)\hat
p_z$.  
Next, we will analyze the effects of the oscillating field coupling term in
numerical simulations and show that an undesired explosion of the
wavefunction can be avoided by using the standard untransformed coupling 
even in the absorption region.   

\subsection{ECS technique in the length and the velocity gauges}

In order to investigate the effects of the oscillating field on the transformed
wavefunction we have performed 1D test calculations with a soft atomic model
potential 
\begin{equation}
V_0 = 
-\frac{1}{\sqrt{1+z^2}}
\end{equation}
using two different external fields, namely a high-frequency electric field 
given by
\begin{eqnarray}
E=
\left\{\begin{array}{ll} 
E_1\sin(\omega_1 t) t/5T,& \textrm {as $t \leq 5T$}\\
E_1\sin\omega_1 t, &\textrm {as $t> 5T$}
\end{array}\right.
\label{field1}
\end{eqnarray}
where $E_1=0.5$, $\omega_1=0.5$ and $T=2\pi/\omega_1$, and a 3-cycle
low-frequency Ti:sapphire laser pulse 
\begin{equation}
E=E_2\cos(\omega_2 t)\sin^2(\pi t/L_2)\label{Ti}
\label{field2}
\end{equation}
with $E_2 = 0.1$ a.u., $\omega_2=0.057$ a.u. and $L=330$ a.u.. 
The respective vector potentials are derived from the electric field
expressions in Eqs. (\ref{field1}) and (\ref{field2}).
Please note that the former field, which is smoothly turned on over 5 optical
cycles, is similar to the field form used by McCurdy et
al. \cite{McCurdy912}. The real part of the calculation box is restricted in
both cases by $z_1 = -z_2 = -25$ a.u., with the complex part extending over 
12.5 a.u. on both sides of the grid.

\begin{figure}
\includegraphics[width=0.95\columnwidth]{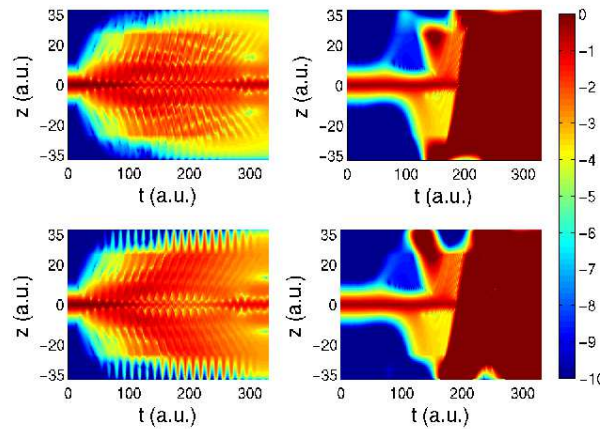}
\caption{\label{full}
(color online) 
Temporal evolution of the probability density from the numerical simulation of
a 1D model atom interacting with a high-frequency field (left hand column) and
a Ti:sapphire field (right hand column). Shown is a comparison of the results
obtained using the external field coupling in the velocity gauge (upper row)
and in the length gauge (lower row).
}
\end{figure}

\begin{figure}
\includegraphics[width=0.95\columnwidth]{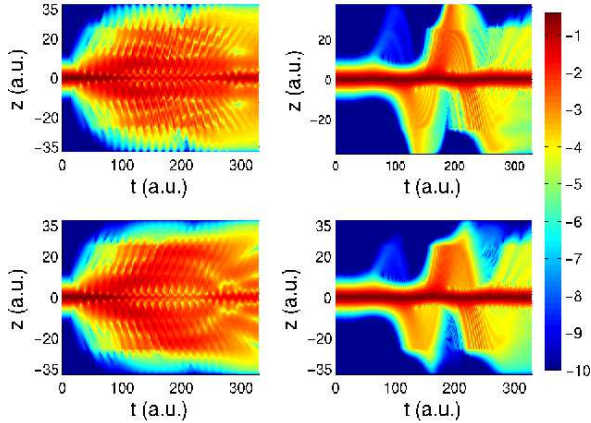}
\caption{\label{part}
(color online) 
Same as Fig. \ref{full} but using the untransformed standard field coupling on
the entire grid.
}
\end{figure}

In Fig. \ref{full}  we present the temporal 
evolution of the electron density distributions in the high-frequency field
(left hand column) and in the Ti:sapphire field (right hand column). 
The panels in the upper and the lower row show the numerical results obtained
in the length and the velocity gauge, respectively. 
The effect of the first factor in Eqs. (\ref{ecs1}) and (\ref{ecs2}) is most
clearly seen in the results of the low-frequency calculations, where we observe 
an explosion of the wave-funtion in the second part of the evolution after a 
significant part of the wavefunction has entered the complex area. This is
obviously due to the fact that the term acts as a source over a half cycle of
the pulse. As expected above the same unphysical feature is found in the
length as well as in the velocity gauge. 
In the high-frequency case (left-hand column) the results obtained in the two
gauges are again similar, but we do not observe a significant amount of
explosion. This difference as compared to the Ti:sapphire calculations is 
probably due to two factors: First, in the high-frequency simulation the
probability density, which enters the complex area, is smaller than in the 
low-frequency calculations. Second, the rapid change of the sign of the 
electric field or the vector potential may effectively prevent an explosion, 
since the complex factor quickly changes between decay and source nature.

From the results presented above we may therefore infer that the ECS
transformation may lead to unphysical results in both gauges due to the
oscillating nature of the external field.  
The most straightforward strategy to circumvent this problem appears to
simply neglect the ECS transformation of the coordinates in the field 
coupling term. This means in other words that the untransformed standard 
field coupling (c.f. Eq. (\ref{coupling})) is used over the entire grid 
including the absorbing area. 
Using this strategy there is no risk to create a source term while
the desired decay of the wavefunction should be still achieved
via the kinetic operator term. 

In order to test our expectations we have repeated the simulations by dropping
the term $\exp(i\eta)$ in Eq. (\ref{ecs1}) and $\exp(-i\eta)$ in Eq. (\ref{ecs2}), the results are
shown in Fig. \ref{part}. The comparison with the respective panels of
Fig. \ref{part} show immediately that the desired effect is achieved. In
particular in the low-frequency case (right hand column) the wavefunction is
absorbed at the edges of the grid and no signature of explosion is seen 
anymore. 

We may therefore conclude that the exterior complex scaling technique can be
applied as an absorbing boundary to time-dependent simulations on laser-atom
simulations in the length as well as in the velocity gauge 
as long as the complex factors in Eqs. (\ref{ecs1}) and
(\ref{ecs2}) are removed. We may note that McCurdy et al. 
\cite{McCurdy912} reached to a similar conclusion for the velocity gauge as 
they did not transform the momentum $\hat p_Z$ to the complex plane. 
Our analysis above shows that an analogous restriction is possible in the 
length gauge as well. 
Our test calculations have shown that, in general, after
omission of the unstable factors calculations in the length gauge show
slightly better results than those in the velocity gauge. We therefore
restrict ourselves below to the length gauge only.

Thus, the above implementation of the ECS technique
coincides with the desired absorbing boundary condition 
in time-dependent strong-field calculations.
It has been shown in the application of ECS to the time-independent
Schr\"odinger equation (e.g. \cite{Rescigno99,McCurdy04b} and for review
\cite{McCurdy04}) that using a
sharp exterior scaling the derivative discontinuity at the boundary
is handled exactly, as long as the boundary is chosen to coincide with
a grid point.
We have adapted this strategy in the time-dependent calculations.
In test calculations
we have found that the absorption effect at the boundaries
is almost independent of the scaling angle $\eta$, in the present calculations
we have used $\eta=\pi/3$.
Finally,
before proceeding with a comparison of the ECS results with those
  obtained using the standard masking function technique 
we may note that special care
has to be taken in the representation of the first and second
derivatives in the transformed Schr\"odinger equation
(for a detailed discussion, see \cite{McCurdy04}), which
we have approximated using Lagrange interpolating polynomials \cite{http}.

\subsection{Comparison of ECS and masking function techniques as absorber}

In order to analyze the efficiency of the ECS method as absorber in
time-dependent simulations as compared to the masking function technique
we present in this section results of
calculations for the interaction of 
the hydrogen atom with a linearly polarized laser
field. We compare the results obtained using the ECS method with those obtained
using conventional masking functions of the form:
\begin{equation}
\label{mask}
M = \cos^{1\over 8}\left({|X-x_i|\over d}{\pi \over 2}\right),
\end{equation}
where $d$ is the length of the absorbing region, over which $M$ changes
smoothly from 1 to 0, and $x_i$ is the boundary point.
Such a function has been applied at all boundaries of the numerical grid
(c.f. gray zones in Fig. \ref{f1}).
In the course of the calculations we have tested different
masking functions, the results presented
below are found to be rather insensitive on the form of masking functions.

We use the time-dependent Schr\"odinger equation in the length gauge
and the Coulomb potential of the hydrogen atom,
\begin{equation}
V(z,\rho) = E(t)z - \frac{1}{\sqrt{z^2 + \rho^2}}.
\end{equation}
The electric field profile is given by Eq. (\ref{Ti}). 
The field parameters are the same as before.
The grid parameters are $\Delta z = 0.1$ a.u.,
$\Delta\rho = 0.2$ a.u. and the time step $\Delta t = 0.1$ a.u..
The initial ground states have been obtained via imaginary time propagation
\cite{Kosloff86}. The absorber is applied at $\rho_0=22$ and $z_1=-z_2=-10$. 
The width of the absorber is chosen to be 20\% of the grid size.

\begin{figure}
\includegraphics[width=1.0\columnwidth]{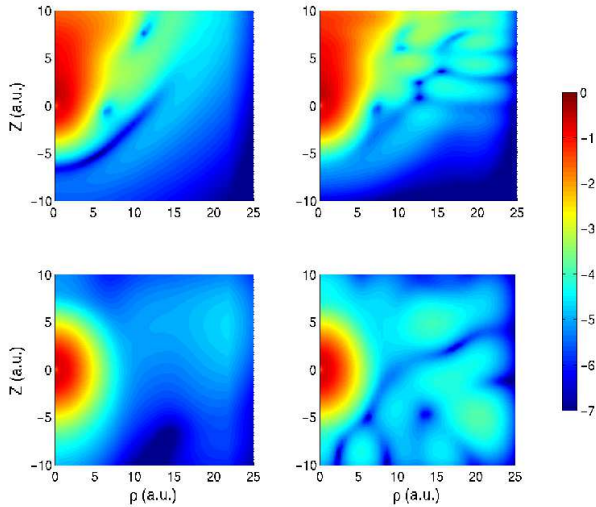}
\caption{ (color online) Logarithmic contour plots of the electron
density distributions from the numerical solution of the 2D TDSE
of the hydrogen atom exposed to an intense laser pulse at $t=165$ a.u.
(upper row) and $t=330$ a.u. (lower row). Results obtained using ECS
(left hand panels) are compared with those obtained using the
masking function (right hand panels) as absorber. Field parameters
as in Eq. (\ref{Ti}). } \label{2D}
\end{figure}

To demonstrate that the reduction of reflections is present
in the solution of the TDSE,
we present in Fig. \ref{2D} probability density distributions.
From the comparison between the results for
ECS (left column) and for the masking function (right column) at $t=165$ a.u. and
$t=330$ a.u. in the upper and lower row, respectively,
the difference in the efficiency of the two absorbers
is clearly visible.
The distributions obtained with the masking function show
interference patterns due to the reflections at the boundaries, which
are not seen in the results for the ECS absorber.
Note that at $t=165$ a.u.
the wave packet has reached (and is reflected from) the upper
boundary in $Z$-direction only,
while at the end of the pulse reflections in all
directions have appeared.

\section{Calculation of high harmonic spectra}

Finally, we apply the ECS absorber to a typical intense-field
phenomenon, namely the evaluation of high harmonic spectra. High
harmonic generation (HHG) is an important process for laser
frequency conversion and the generation of attosecond pulses (for
reviews, see e.g. \cite{Brabec00,Agostini04}). According to the
semiclassical three-step rescattering picture
\cite{Schafer93,Corkum93}, and confirmed by quantum-mechanical
calculations \cite{Lewenstein94}, HHG can be understood as the
ionization of an electron by tunneling through the barrier of the
combined Coulomb and laser fields, followed by the acceleration of
the electron in the field, which may cause, for linear
polarization of the field, a return of the electron to and its
recombination with the parent ion under the emission of a harmonic
photon. On the basis of this picture it is reasonable to limit the
grid size of an ab-initio calculation of high harmonic spectra via
the time-dependent Schr\"odinger equation, since beyond a certain
distance from the nucleus outgoing wave packets are expected to
have no effect on the high harmonic spectra.

\begin{figure}
\includegraphics[width=1\columnwidth]{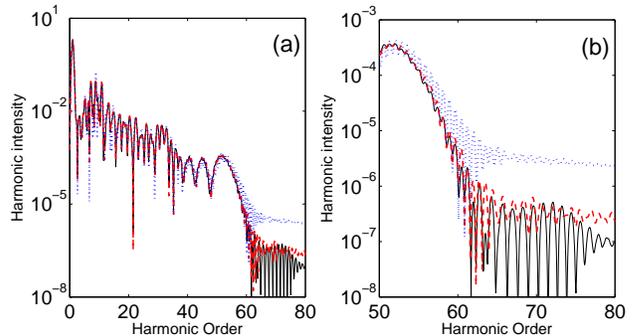}
\caption{\label{f4} (color online)
High harmonic spectra from the interaction of the hydrogen atom with an intense
linearly polarized 3-cycle laser pulse at $E_0 = 0.1$ a.u. and $\omega = 0.057$ a.u..
Shown is a comparison between the results obtained from the (unrestricted)
reference calculation (black solid line) and calculations using the ECS
(red dashed line) and the masking function (blue dotted line) as absorbers
at $z_1 = -z_2 = -80$ a.u.. In transversal direction the grid was large enough to
avoid additional reflections from this boundary.
(a) Full spectrum and (b) cut-off region.
}
\label{HHG1}
\end{figure}

We have performed simulations of the 2D TDSE for the hydrogen atom in an
intense linearly polarized laser pulse, given by the pulse form in
Eq. (\ref{field2}) with $E_0 =0.1$ a.u. and $\omega = 0.057$ a.u., and evaluated
high harmonic spectra as the Fourier transform of the time-dependent dipole
moment. 
Note that the dipole moment has been determined over the interior box
(i.e. without the absorbing regions).
In order to analyze the effect of reflections
from the edges of the 
grid along the polarization direction on the spectra, 
 we compare in Fig. \ref{HHG1} the results of three simulations,
namely the full calculation as a reference (black solid line),
performed on a sufficiently large simulation box
to prevent reflections at the boundaries,
and calculations using ECS (red dashed line) and
masking function (blue dotted line) as absorbers.
The calculations with absorbers have been restricted along the polarization
axis by choosing $z_1=-z_2=-80$ a.u., which
exceeds the maximum excursion of the classical electron trajectories of 63.6
a.u..
The absorbing part of the grid has been chosen to extend over an
additional 20 a.u. at both ends. In the transversal direction the grid was
chosen large enough to avoid reflections from this boundary.

The results in Fig. \ref{HHG1}a) show the typical high harmonic spectrum with
a plateau and a cut-off at $N = (I_p + 3.17 U_p)/\omega \approx 51$, where
$I_p = 0.5$ is the ionization potential of the hydrogen atom and
$U_p = I_0/4\omega^2 = 0.77$ is the ponderomotive potential.
From the comparison in Fig. \ref{HHG1} it is seen that
the harmonics in the plateau do not differ significantly.
There are small deviations in the
minima between the harmonics obtained
from the simulation with the masking function
but the maxima appear to be unchanged. The effects of the reflections become
visible at and beyond the cut-off, this region is enlarged in
Fig. \ref{HHG1}b). While the results from the ECS calculation
almost agree with those from the full calculation over a decrease in the signal
of two orders of magnitude, the results evaluated with the masking function
start to differ near the cut-off and the deviations increase up to an order
of magnitude in the signal beyond the cut-off. 
This ``artificial" increase of the HHG signal
in the cut-off region results from those parts of the wavefunction reflected 
at the boundary, which return to the nucleus and give rise to harmonics
without physical meaning.
Note that an accurate
calculation of the harmonics in the cut-off regime is e.g. important for
analysis of the generation of single attosecond pulses
(e.g. \cite{Hentschel01}).

\begin{figure}
\includegraphics[width=1.0\columnwidth]{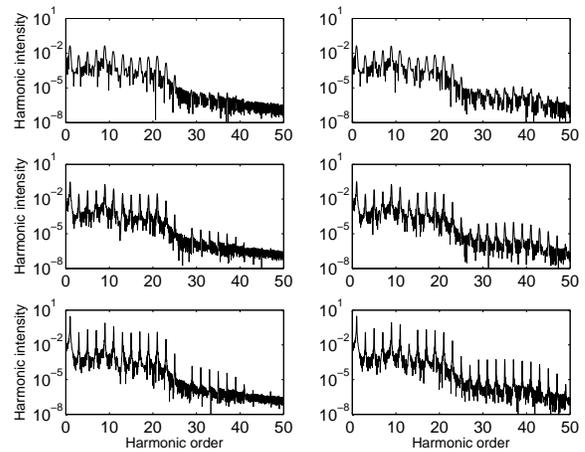}
\caption{\label{f6}
Comparison of harmonic spectra calculated using the ECS (left column) and the
masking function (right column) technique in laser pulse with constant
amplitude having 3 (upper row), 6 (middle row) and 12 (lower row) cycles.
}
\end{figure}

The effect of the reflections even increases for longer pulses,
in which several wave packets reach the boundaries.
This is seen from the results,
presented in Fig. \ref{f6}, where harmonic spectra obtained for laser fields
with a
constant envelope, $E=E_0\sin(\omega t)$ with $E_0 = 0.05$ and $\omega =
0.057$, having 3 (upper row), 6 (middle row) and 12 (bottom row) cycles
are shown. Results obtained using
ECS and masking function as absorber techniques are shown
in the left and right hand panels, respectively.
The absorbers were placed at $z_1=-z_2=-35$ a.u. (here the maximum excursion
length is 30.8 a.u.),
with an additional absorbing region of 6 a.u.
The grid size in $\xi$-direction was big enough not to influence the results.
While the results obtained for the 3-cycle field show the expected cut-off at
the 19th harmonic, spurious harmonics due to reflections
appear beyond the cut-off and increase in
signal as the number of field cycles increases.
The comparison shows that for the ECS results the contrast ratio between
the false and the plateau harmonics is about $10^{-3}$ in all cases,
while for the masking functions it increases, especially for the highest false
harmonics, giving the impression of a second unphysical plateau.

\section{Summary}

In summary, we have investigated the implementation of the
exterior complex scaling technique as an absorber in the numerical
solution of the time-dependent Schr\"odinger equation 
for strong-field problems on a grid. 
Our analysis
has shown that the ECS technique can be applied in both the length and
  the velocity gauge as long as the untransformed field coupling is used on
  the entire grid including the absorbing area. It is found that the decay due
  to the ECS transformation in the kinetic operator term is sufficient to
  efficiently reduce reflections at the grid boundaries.
A comparative study has shown that in this implementation
a significantly better suppression of reflections 
can be achieved as using the conventional masking function
method. 
By application of the ECS method to the evaluation of
high harmonic spectra differences in the suppression of artifacts,
e.g. in form of spurious harmonics, is demonstrated. The simple
test cases considered here should capture the essence of the
reflection problem, exterior complex scaling in both length and velocity
  gauge can be therefore
considered as an efficient absorption technique for numerical
time-dependent solutions in higher dimensions and/or of more
complex processes too.

\section*{Acknowledgment}
We thank S. Baier, P. Panek, L. Plaja, A. Requate, and 
J. R. V\'azquez de Aldana 
for many stimulating discussions. This work has been partially
supported by DAAD via project D/05/25690.

\end{document}